\documentclass[prd,preprint,superscriptaddress,showpacs,byrevtex]{revtex4}
\usepackage{bm}
\def\fsl#1{\setbox0=\hbox{$#1$}           % set a box for #1
   \dimen0=\wd0                                 % and get its size
   \setbox1=\hbox{/} \dimen1=\wd1               % get size of /
   \ifdim\dimen0>\dimen1                        % #1 is bigger
      \rlap{\hbox to \dimen0{\hfil/\hfil}}      % so center / in box
      #1                                        % and print #1
   \else                                        % / is bigger
      \rlap{\hbox to \dimen1{\hfil$#1$\hfil}}   % so center #1
      /                                         % and print /
   \fi}                                         %
\newcommand{\be}{\begin{equation}}
\newcommand{\ee}{\end{equation}}
\newcommand{\bea}{\begin{eqnarray}}
\newcommand{\eea}{\end{eqnarray}}
\newcommand{\beq}{\begin{equation}}
\newcommand{\eeq}{\end{equation}}
\newcommand{\beqs}{\begin{eqnarray}}
\newcommand{\eeqs}{\end{eqnarray}}

\newcommand{\aslash}{A\hspace{-0.067in}\slash}

\begin{document}
\title{ A New Contribution To The Proton Spin Crisis: Non-Zero Boundary Surface Term Due To Confinement in QCD }
\author{Gouranga C Nayak }\thanks{E-Mail: nayakg138@gmail.com}
%
%\affiliation{ C. N. Yang Institute for Theoretical Physics, Stony Brook University, Stony Brook NY, 11794-3840 USA}
%
\date{\today}
\begin{abstract}
The proton spin crisis still remains an unsolved problem in particle physics. It is suggested in the literature that the gauge invariant spin and orbital angular momentum of the quark and the gauge invariant total angular momentum of the gluon contribute to the proton spin. However, in this paper we find a new contribution to the proton spin. This new contribution is due to the boundary surface term in the total angular momentum conservation equation in QCD which is non-zero due to the confinement of the quarks and gluons inside the finite size proton. We derive the non-perturbative formula of this new contribution to the proton spin which can be calculated by using the lattice QCD method.
\end{abstract}
\maketitle
\pagestyle{plain}

\pagenumbering{arabic}

\section{ Introduction }

The proton consists of quarks and gluons. If the proton is in motion at the high energy then it consists of quarks, antiquarks and gluons because sea quarks, antiquarks and gluons are produced from the QCD vacuum. Hence it was expected that the spin of the proton is equal to the sum of the spin of the quarks, antiquarks and gluons inside the proton, {\it i. e.} the spin $\frac{1}{2}$ of the proton was expected to be
\bea
\frac{1}{2} =  S_q+ S_g
\label{hft}
\eea
where $S_q$ is the spin of the quarks plus antiquarks inside the proton and $S_g$ is the spin of the gluons inside the proton. In terms of the spin angular momentum operator we have in eq. (\ref{hft}) [see section II for the definition of the spin/angular momentum operators in terms of the quark and gluon fields in QCD]
\bea
S_q=<p,s|{\hat S}_q|p,s>,~~~~~~~~~~~~~S_g=<p,s|{\hat S}_g|p,s>
\eea
where $|p,s>$ is the longitudinally polarized proton physical energy-momentum eigenstate with momentum $p^\mu$ and spin $s$ and ${\hat S}_q$ is the longitudinal component of the gauge invariant spin angular momentum operator vector ${\hat {\vec S}}_q$ of the quarks plus antiquarks inside the proton and ${\hat S}_g$ is the longitudinal component of the spin angular momentum operator vector ${\hat {\vec S}}_g$ of the gluons inside the proton.

However, the experimental data in late 1980's \cite{emt} found that the sum of the spins of the quarks and antiquarks inside the proton contributes only to a negligible fraction of the proton spin. This was later confirmed by other experiments \cite{oht}. Even the addition of the spins of the gluons inside the proton \cite{hct} cannot explain the spin of the proton. At present it is found that about fifty percent of the proton spin is carried by the spins of the partons inside the proton \cite{wdt}. The remaining spin of the proton is still missing which is known as the proton spin crisis.

In order to explain this remaining proton spin it was suggested that the orbital angular momentum of the quarks and gluons should contribute to the proton spin. In the parton model it was suggested that \cite{jftt}
\bea
\frac{1}{2} = S_q+{\tilde S}_g+{\tilde  L}_q+{\tilde L}_g
\label{jft}
\eea
where ${S}_q$ is the gauge invariant spin of the quarks plus antiquarks inside the proton, ${\tilde  S}_g$ is the spin of the gluons inside the proton in the light-cone gauge, ${\tilde  L}_q$ is the orbital angular momentum of the quarks plus antiquarks inside the proton in the light-cone gauge and ${\tilde  L}_g$ is the orbital angular momentum of the gluons inside the proton in the light-cone gauge. In terms of the spin/orbital angular momentum operator one finds in eq. (\ref{jft})
\bea
S_q=<p,s|{\hat S}_q|p,s>,~~~~{\tilde S}_g=<p,s|{\tilde {\hat S}}_g|p,s>,~~~~~{\tilde L}_q=<p,s|{\tilde {\hat L}}_q|p,s>,~~~~~{\tilde L}_g=<p,s|{\tilde {\hat L}}_g|p,s>\nonumber \\
\label{jfta}
\eea
where ${\hat S}_q$ is the longitudinal component of the gauge invariant spin angular momentum operator vector ${\hat {\vec S}}_q$ of the quarks plus antiquarks inside the proton, ${\tilde {\hat S}}_g$ is the longitudinal component of the spin angular momentum operator vector ${\tilde {\hat {\vec S}}}_g$ of the gluons inside the proton in the light-cone gauge, ${\tilde {\hat L}}_q$ is the longitudinal component of the gauge invariant spin angular momentum operator vector ${\tilde {\hat {\vec L}}}_q$ of the quarks plus antiquarks inside the proton in the light-cone gauge and ${\tilde {\hat S}}_g$ is the longitudinal component of the spin angular momentum operator vector ${\tilde {\hat {\vec S}}}_g$ of the gluons inside the proton in the light-cone gauge.

In terms of the gauge invariant definition of the spin/angular momentum one finds \cite{xft}
\bea
\frac{1}{2} =  S_q+ L_q+ J_g
\label{xjft}
\eea
where $S_q$ is the gauge invariant spin of the quarks plus antiquarks inside the proton, $L_q$ is the gauge invariant orbital angular momentum of the quarks plus antiquarks inside the proton and $J_g$ is the gauge invariant total angular momentum of the gluons inside the proton. In terms of the gauge invariant angular momentum operator one finds in eq. (\ref{xjft})
\bea
S_q= <p,s|{\hat S}_q|p,s>,~~~~~~~~~~~~~~~~~L_q=<p,s|{ {\hat L}}_q|p,s>,~~~~~~~~~~~~~~~J_g=<p,s|{\hat J}_g|p,s>
\label{xjfta}
\eea
where ${\hat S}_q$ is the longitudinal component of the gauge invariant spin angular momentum operator vector ${\hat {\vec S}}_q$ of the quarks plus antiquarks inside the proton, ${\hat L}_q$ is the longitudinal component of the gauge invariant orbital angular momentum operator vector ${\hat {\vec L}}_q$ of the quarks plus antiquarks inside the proton and ${\hat J}_g$ is the longitudinal component of the gauge invariant total angular momentum operator vector ${\hat {\vec J}}_g$ of the gluons inside the proton.

However, it is recently shown that the eq. (\ref{xjft}) is not consistent with the conservation of angular momentum in QCD if the confinement effect is taken into account. This is because the angular momentum sum rule is violated in QCD \cite{nkagt} due to the non-zero boundary surface term in the angular momentum conservation equation in QCD \cite{nkymt} due to the confinement of quarks and gluons inside the finite size hadron \cite{nkbst}.

Hence when the confinement effect in QCD is taken into account we find \cite{nkagt}
\bea
\frac{1}{2} = S_q+L_q+J_g+J_S
\label{nkft}
\eea
where the gauge invariant definition of the $S_q$, $L_q$, $J_g$ are given by eq. (\ref{xjfta}) and $J_S$ is the boundary surface term contribution in the total angular momentum conservation equation in QCD which is non-zero due to the confinement of quarks and gluons inside the finite size hadron \cite{nkbst}. The gauge invariant definition of the boundary surface term $J_S$ is given by
\bea
J_S=<p,s|{\hat J}_S|p,s>
\label{nkfta}
\eea
where ${ J}_S$ is the longitudinal component of the gauge invariant vector ${\vec J}_S$ as given by eq. (\ref{jqgt}).

Therefore we find from eq. (\ref{nkft}) that the angular momentum sum rule as given by eq. (\ref{xjft}) is violated in QCD due to the confinement of quarks and gluons inside the finite size proton \cite{nkbst}).

We find from eq. (\ref{nkft}) that there is a new contribution to the proton spin which is $J_S$ as given by eq. (\ref{nkfta}) where ${\vec J}_S$ is given by eq. (\ref{jqgt}). This new contribution $J_S$ in eq. (\ref{nkfta}) is a non-perturbative quantity in QCD which cannot be calculated by using the perturbative QCD (pQCD). On the other hand the analytical solution of the non-perturbative QCD is not known. Hence the lattice QCD method can be used to calculate this non-perturbative quantity $J_S$ in eq. (\ref{nkfta}).

In this paper we derive the non-perturbative formula of the $J_S$ in eq. (\ref{nkfta}) from the first principle which can be calculated by using lattice QCD method. This non-perturbative formula of $J_S$ is derived in eq. (\ref{v3c}) which can be calculated by the using lattice QCD method.

We will present a derivation of eq. (\ref{v3c}) in this paper.

The paper is organized as follows. In section II we discuss the angular momentum sum rule violation in QCD due to confinement of quarks and gluons inside finite size hadron. In section III we derive the non-perturbative formula of this new contribution $J_S$ to the proton spin which can be calculated by using the lattice QCD. Section IV contains conclusions.

\section{ Angular momentum sum rule violation in QCD due to confinement }

From the gauge invariant Noether's theorem in QCD we find the continuity equation \cite{nkymt}
\bea
\partial_\mu J^{\mu \nu \lambda}(x) =0
\label{jmt}
\eea
where the gauge invariant angular momentum tensor density $J^{\mu \nu \lambda}(x)$ in QCD is given by
\bea
J^{\mu \nu \lambda}(x) = T^{\mu \lambda}(x) x^\nu -T^{\mu \nu}(x) x^\lambda.
\label{jmnt}
\eea
In eq. (\ref{jmnt}) the $T^{\mu \nu}(x)$ is the gauge invariant and symmetric energy-momentum tensor density in QCD given by
\bea
&& T^{\mu \nu}(x) = F^{\mu \lambda d}(x)F_\lambda^{~ \nu d}(x)+\frac{1}{4} g^{\mu \nu} F^{\lambda \delta d}(x) F_{\lambda \delta}^d(x) +\frac{i}{4} {\bar \psi}_j(x)[\gamma^\mu (\delta^{jk} {\overrightarrow \partial}^\nu-igT^d_{jk}A^{\nu d}(x)) \nonumber \\
&&+\gamma^\nu (\delta^{jk} {\overrightarrow \partial}^\mu-igT^d_{jk}A^{\mu d}(x))-\gamma^\mu (\delta^{jk} {\overleftarrow \partial}^\nu+igT^d_{jk}A^{\nu d}(x)) -\gamma^\nu (\delta^{jk} {\overleftarrow \partial}^\mu+igT^d_{jk}A^{\mu d}(x))]\psi_k(x)\nonumber \\
\label{tmnt}
\eea
where $\psi_i(x)$ is the quark field with color index $i=1,2,3$, the $A_\nu^b(x)$ is the gluon field with color index $b=1,...,8$, the Lorentz index $\nu=0,1,2,3$ and
\bea
F_{\nu \lambda}^c(x)=\partial_\nu A_\lambda^c(x) - \partial_\lambda A_\nu^c(x)+gf^{cda} A_\nu^d(x) A_\lambda^a(x).
\label{fmt}
\eea

From eq. (\ref{jmt}) we find
\bea
\frac{d}{dt} <p,s|[{\vec {\hat S}}_q(t)+{ \vec {\hat L}}_q(t) +{ \vec {\hat J}}_g(t)+{ \vec {\hat J}}_S(t)]|p,s>=0
\label{cet}
\eea
where
\bea
{ \vec {\hat S}}_q(t) = \int d^3x ~\psi_j^\dagger(x) {\vec \sigma} \psi_j(x),
\label{sqt}
\eea
\bea
{ \vec {\hat L}}_q(t) = \int d^3x~ {\vec x} \times \psi_j^\dagger(x) [-i{\overrightarrow D}_{jk}[A]+i{\overleftarrow D}_{jk}[A]] \psi_k(x),
\label{lqt}
\eea
\bea
{ \vec {\hat J}}_g(t) = \int d^3x~ {\vec x} \times {\vec E}^c(x) \times {\vec B}^c(x)
\label{lgt}
\eea
and
\bea
&& {\hat J}^k_S(t) = \epsilon^{kjl} \int d^4x~\partial_n [x^j[E^{lc}(x)E^{nc}(x)-\delta^{nl}\frac{{E_c}^2(x)-{B_c}^2(x)}{2}] \nonumber \\
&&+[x^j\frac{i}{2}{\bar \psi}_{n'}(x)[\gamma^n(\delta_{n'n''}{\overrightarrow \partial}^l-igT^c_{n'n''}A^{lc}(x))-\gamma^n(\delta_{n'n''}{\overleftarrow \partial}^l+igT^c_{n'n''}A^{lc}(x))]\psi_{n''}(x)+\frac{1}{8}{\bar \psi}_{n'}(x)\{\gamma^n,\sigma^{jl}\}]].\nonumber  \\
\label{jqgt}
\eea
In eq. (\ref{sqt}) the $\sigma^i$ is the Pauli spin matrix, in eq. (\ref{lqt}) the covariant derivative $D_{ij}[A]$ in the fundamental representation of SU(3) is given by
\bea
D_{n'n''}[A]=\delta_{n'n''}(i{\not \partial}-m)-igT^c_{n'n''}\aslash^c(x)
\eea
and in eq. (\ref{jqgt}) the Dirac tensor $\sigma^{\mu \nu}$ is given by
\bea
\sigma^{jl}=\frac{i}{2}[\gamma^j,\gamma^l].
\eea
Note that the sum over all the quarks, antiquarks and gluons inside the proton is understood in eq. (\ref{cet}).

From eqs. (\ref{sqt}), (\ref{lqt}), (\ref{lgt}), (\ref{jqgt}) and (\ref{cet}) we find
\bea
\frac{d}{dt}[S_q+ L_q+ J_g]=-\frac{dJ_S}{dt}
\label{xjftxa}
\eea
where
\bea
S_q= <p,s|{{\hat S}}^z_q|p,s>,~~~~L_q=<p,s|{ {\hat L}}^z_q|p,s>,~~~~J_g=<p,s|{{\hat J}}^z_g|p,s>,~~~~J_S=<p,s|{{\hat J}}^z_S|p,s>\nonumber \\
\label{xjfty}
\eea
where the superscript $z$ means the z-component (the longitudinal component) of the vector and the suppression of the $t$ dependence is understood.

Since the boundary surface term is non-zero due to the confinement of quarks and gluons inside the finite size proton \cite{nkbst} we find that \cite{nkagt}
\bea
\frac{dJ_S}{dt}\neq 0.
\label{neq1}
\eea
From eqs. (\ref{neq1}) and (\ref{xjftxa}) we find
\bea
\frac{d}{dt}[S_q+ L_q+ J_g]\neq 0
\label{xjftx}
\eea
which implies that $S_q+ L_q+ J_g$ is time dependent.

Since the proton spin $\frac{1}{2}$ is time independent and the $S_q+ L_q+ J_g$ from eq. (\ref{xjftx}) is time dependent we find that
\bea
\frac{1}{2} \neq S_q+ L_q+ J_g
\label{neq}
\eea
which does not agree with eq. (\ref{xjft}).

This implies that the angular momentum sum rule in QCD as given by eq. (\ref{xjft}) is violated due to the confinement of the quarks and gluons inside the finite size hadron.

\section{Non-Perturbative angular momentum boundary surface Term contribution to the proton spin using lattice QCD }

From eq. (\ref{xjftxa}) we find
\bea
\frac{d}{dt}[S_q+ L_q+ J_g+J_S]=0
\label{xjftxbc}
\eea
which implies that $S_q+ L_q+ J_g+J_S$ is time independent.
Since the proton spin $\frac{1}{2}$ is time independent and the $S_q+ L_q+ J_g+J_S$ from eq. (\ref{xjftxbc}) is time independent we find, unlike eq. (\ref{neq}), that
\bea
\frac{1}{2} = S_q+L_q+J_g+J_S
\label{nkftxa}
\eea
which agrees with eq. (\ref{nkft}).

Hence we find that the new contribution to the proton spin is $J_S$ where $J_S$ is given by eq. (\ref{xjfty}) with ${{\hat J}}^z_S$ given by eq. (\ref{jqgt}).
This new contribution $J_S$ is a non-perturbative quantity in QCD which cannot be calculated by using the perturbative QCD (pQCD). On the other hand the analytical solution of the non-perturbative QCD is not known. Hence the lattice QCD method can be used to calculate this non-perturbative quantity $J_S$. Recently there has been new development in lattice QCD method to study various non-perturbative quantities in QCD in vacuum \cite{nkalt} and in QCD in medium \cite{nkalt1} to study quark-gluon plasma at RHIC and LHC \cite{qgt,qgt1,qgt2,qgt3}. In this section we derive the non-perturbative formula of the $J_S$ from the first principle which can be calculated by using lattice QCD method.

The partonic operator to form proton is given by
\bea
{\hat {\cal O}}_p(x) =\epsilon_{kln}\psi_u^k(x) C\gamma^5 \psi_d^l(x) \psi_u^n(x)
\label{prt}
\eea
where $u,d$ refers to up, down quarks, $C$ is the charge conjugation operator and $k,l,n=1,2,3$ are color indices. In order to calculate $J_S$ we need to study the ratio of the three-point non-perturbative correlation function in QCD
\bea
C_3^k(p)=\sum_{r'} <\Omega|e^{-i{\vec p}\cdot {\vec r}'} {\hat {\cal O}}_p(t',r') {\hat J}^k_S(t) {\hat {\cal O}}_p(0)|\Omega>
\label{c3ct}
\eea
to the two-point non-perturbative correlation function in QCD
\bea
C_2(p)=\sum_{r'} <\Omega|e^{-i{\vec p}\cdot {\vec r}'} {\hat {\cal O}}_p(t',r') {\hat {\cal O}}_p(0)|\Omega>
\label{c2ct}
\eea
where the operator ${\hat J}^k_S$ is given by eq. (\ref{jqgt}), the operator ${\hat {\cal O}}_p(t,r)$ is given by eq. (\ref{prt}) and $|\Omega>$ is the vacuum state of the full QCD (not the vacuum state of the pQCD).

In eq. (\ref{c3ct}) the vacuum expectation of the non-perturbative three-point correlation function in QCD is given by
\bea
&&<\Omega|{\hat {\cal O}}_p(t',r'){\hat J}^k_S(t'')  {\hat {\cal O}}_p(0)|\Omega>=\frac{1}{Z[0]} \int [d{\bar \psi}_u][d{ \psi}_u] [d{\bar \psi}_d][d{ \psi}_d][dA]~{\hat {\cal O}}_p(t',r') {\hat J}^k_S(t'') {\hat {\cal O}}_p(0) \times {\rm det}[\frac{\delta G^c_f}{\delta \omega^d}] \nonumber \\
&& \times e^{i\int d^4x [-\frac{1}{4} F_{\nu \lambda}^b(x)F^{\nu \lambda b}(x) -\frac{1}{2\alpha} [G_f^b(x)]^2 +{\bar \psi}^u_j(x)[\delta^{jk}(i{\not \partial}-m_u)+gT^b_{jk}\aslash^b(x)]\psi^u_k(x) +{\bar \psi}^d_j(x)[\delta^{jk}(i{\not \partial}-m_d)+gT^b_{jk}\aslash^b(x)]\psi^d_k(x)]} \nonumber \\
\label{cf3t}
\eea
and in eq. (\ref{c2ct}) the vacuum expectation of the non-perturbative two-point correlation function in QCD is given by
\bea
&&<\Omega|{\hat {\cal O}}_p(t',r') {\hat {\cal O}}_p(0)|\Omega>=\frac{1}{Z[0]} \int [d{\bar \psi}_u][d{ \psi}_u] [d{\bar \psi}_d][d{ \psi}_d][dA]~{\hat {\cal O}}_p(t',r') {\hat {\cal O}}_p(0) \times {\rm det}[\frac{\delta G^c_f}{\delta \omega^d}] \nonumber \\
&& \times e^{i\int d^4x [-\frac{1}{4} F_{\nu \lambda}^b(x)F^{\nu \lambda b}(x) -\frac{1}{2\alpha} [G_f^b(x)]^2 +{\bar \psi}^u_j(x)[\delta^{jk}(i{\not \partial}-m_u)+gT^b_{jk}\aslash^b(x)]\psi^u_k(x) +{\bar \psi}^d_j(x)[\delta^{jk}(i{\not \partial}-m_d)+gT^b_{jk}\aslash^b(x)]\psi^d_k(x)]} \nonumber \\
\label{cf2t}
\eea
where $G_f^a(x)$ is the gauge fixing term, $\alpha$ is the gauge fixing parameter and
\bea
&&Z[0]= \int [d{\bar \psi}_u][d{ \psi}_u] [d{\bar \psi}_d][d{ \psi}_d][dA]~{\hat {\cal O}}_p(t,r) {\hat {\cal O}}_p(0) \times {\rm det}[\frac{\delta G^c_f}{\delta \omega^d}] \nonumber \\
&& \times e^{i\int d^4x [-\frac{1}{4} F_{\nu \lambda}^b(x)F^{\nu \lambda b}(x) -\frac{1}{2\alpha} [G_f^b(x)]^2 +{\bar \psi}^u_j(x)[\delta^{jk}(i{\not \partial}-m_u)+gT^b_{jk}\aslash^b(x)]\psi^u_k(x) +{\bar \psi}^d_j(x)[\delta^{jk}(i{\not \partial}-m_d)+gT^b_{jk}\aslash^b(x)]\psi^d_k(x)]} \nonumber \\
\label{zt}
\eea
is the generating functional in QCD.

In order to calculate the $J_S$ we need to evaluate the ratio
\bea
\frac{C^k_3(p)}{C_2(p)}=\frac{\sum_{r'} <\Omega|e^{-i{\vec p}\cdot {\vec r}'} {\hat {\cal O}}_p(t',r') {\hat J}^k_S(t) {\hat {\cal O}}_p(0)|\Omega>}{\sum_{r'} <\Omega|e^{-i{\vec p}\cdot {\vec r}'} {\hat {\cal O}}_p(t',r') {\hat {\cal O}}_p(0)|\Omega>}.
\label{r3ct}
\eea
Inserting complete set of proton energy eigenstates
\bea
\sum_l |p_n,s><p_n,s|=1
\label{csct}
\eea
we find in the Euclidean time
\bea
&&\frac{C^k_3(p)}{C_2(p)}=\frac{1}{\sum_n < \Omega|{\hat {\cal O}}_p(0)|p_l,s><p_l,s| {\hat {\cal O}}_p(0)|\Omega>e^{-\int_0^{t'}dt'' E_l(t'')}}\nonumber \\
&&\times ~[\sum_l \sum_{l'} <\Omega|{\hat {\cal O}}_p(0)|p_l,s><p_l,s| {\hat J}^k_S(t) |p_{l'},s><p_{l'},s|{\hat {\cal O}}_p(0)|\Omega>\times ~e^{-\int_0^{t'}dt'' E_l(t'')}]\nonumber \\
\label{s3ct}
\eea
where we have used
\bea
H|p_l,s>=E_{l}(t)|p_l,s>.
\label{cegt}
\eea

Neglecting all the higher energy level contribution in the large time limit $t' \rightarrow \infty $ and keeping only the ground state contribution we find from eq. (\ref{s3ct})
\bea
&&[\frac{C^k_3(p)}{C_2(p)}]_{t'\rightarrow \infty} =\frac{  |<\Omega|{\hat {\cal O}}_p(0)|p,s>|^2<p,s| {\hat J}^k_S(t) |p,s>e^{-\int_0^{t'}dt'' E(t'')}}{|< \Omega|{\hat {\cal O}}_p(0)|p,s>|^2e^{-\int_0^{t'}dt'' E(t'')}}
\label{t3ct}
\eea
which gives
\bea
&&[\frac{C^k_3(p)}{C_2(p)}]_{t'\rightarrow \infty} =<p,s| {\hat J}^k_S(t) |p,s>
\label{u3ct}
\eea
where $|p,s>=|p_0,s>$ is the ground state energy-momentum eigenstate of the proton.

From eqs. (\ref{u3ct}), (\ref{r3ct}) and (\ref{xjfty}) we find
\bea
J_S= [\frac{<\Omega|\sum_{{\vec x}'}e^{-i{\vec p}\cdot {\vec x}'}~{\hat {\cal O}}_p({\vec x}',t') {\hat J}^{k=3}_S(t) {\hat {\cal O}}_p(0)|\Omega>}{<\Omega|\sum_{{\vec x}'}e^{-i{\vec p}\cdot {\vec x}'}~{\hat {\cal O}}_p({\vec x}',t') {\hat {\cal O}}_p(0)|\Omega>}]_{t'\rightarrow \infty}
\label{v3c}
\eea
where ${\vec p}$ is the momentum of the proton, $|\Omega>$ is the vacuum state of the full QCD (not pQCD), ${\hat {\cal O}}_p(x)$ is given by eq. (\ref{prt}) and ${\hat J}^{k}_S(t)$ is given by eq. (\ref{jqgt}).

Since the vacuum expectation of the non-perturbative three-point correlation function $<\Omega|{\hat {\cal O}}_p({\vec x}',t') {\hat J}^{k}_S(t) {\hat {\cal O}}_p(0)|\Omega>$ in QCD can be calculated by using the lattice QCD method using eq. (\ref{cf3t}) and the vacuum expectation of the non-perturbative two-point correlation function $<\Omega|{\hat {\cal O}}_p({\vec x}',t') {\hat {\cal O}}_p(0)|\Omega>$ in QCD can be calculated by using the lattice QCD method using eq. (\ref{cf2t}), we find that the new contribution $J_S$ to the proton spin in eq. (\ref{nkft}) can be calculated by using the lattice QCD method from eq. (\ref{v3c}).

\section{Conclusions}
The proton spin crisis still remains an unsolved problem in particle physics. It is suggested in the literature that the gauge invariant spin and orbital angular momentum of the quark and the gauge invariant total angular momentum of the gluon contribute to the proton spin. However, in this paper we have found a new contribution to the proton spin. This new contribution is due to the boundary surface term in the total angular momentum conservation equation in QCD which is non-zero due to the confinement of the quarks and gluons inside the finite size proton. We have derived the non-perturbative formula of this new contribution to the proton spin which can be calculated by using the lattice QCD method.

\end{document}